\RequirePackage{fix-cm}
\documentclass[twocolumn]{svjour3}          
\smartqed  
\usepackage{graphicx}
\usepackage{natbib}
\usepackage{hyperref}
\usepackage{caption}
\usepackage{amssymb}
\usepackage{multirow}
\usepackage{subfigure}
\usepackage{booktabs}

\begin{document}

\title{Online Learning for Social Spammer Detection on Twitter
}


\author{Phuc Tri Nguyen         \and
        Hideaki Takeda 
}


\institute{Phuc Tri Nguyen \at
            University of Information Technology,  \\
            Ho Chi Minh, Vietnam  \\
            \email{phucnt@uit.edu.vn}           
           \and
           Hideaki Takeda \at
            National Institute of Informatics \\
            Tokyo, Japan\\
            \email{takeda@nii.ac.jp} 
}

\date{Received: date / Accepted: date}

\maketitle

\begin{abstract}
Social networking services like Twitter have been playing an import role in people's daily life since it supports new ways of communicating effectively and sharing information. The advantages of these social network services enable them rapidly growing. However, the rise of social network services is leading to the increase of unwanted, disruptive information from spammers, malware discriminators, and other content polluters. Negative effects of social spammers do not only annoy users, but also lead to financial loss and privacy issues. There are two main challenges of spammer detection on Twitter. Firstly, the data of social network scale with a huge volume of streaming social data. Secondly, spammers continually change their spamming strategy such as changing content patterns or trying to gain social influence, disguise themselves as far as possible. With those challenges, it is hard to directly apply traditional batch learning methods to quickly adapt newly spamming pattern in the high-volume and real-time social media data. We need an anti-spammer system to be able to adjust the learning model when getting a label feedback. Moreover, the data on social media may be unbounded. Then, the system must allow update efficiency model in both computation and memory requirements. Online learning is an ideal solution for this problem. These methods incrementally adapt the learning model with every single feedback and adjust to the changing patterns of spammers overtime. Our experiments demonstrate that an anti-spam system based on online learning approach is efficient in fast changing of spammers comparing with batch learning methods. We also attempt to find the optimal online learning method and study the effectiveness of various feature sets on these online learning methods.
\end{abstract}

\section{Introduction}
\label{intro}
Social network services like Twitter have been played a major role in people's daily life. It offers a convenient and efficient way to communicate and disseminate information. Individuals use Twitter to tweet anything about their concern such as news, jokes, or even their feeling. Companies and organizations use Twitter as an effective channel to connect with their customers, promote or sell their products. With these advantages, Twitter has been increasingly used for large-scale information dissemination in various fields of human life such as marketing, journalism or public relations.\\\\
Nevertheless, the popularity of Twitter has led to the rise of unwanted, disruptive information from social spammers. Twitter spammers \citep{wiki:spamming} are defined as malicious users who try to gain social influence and generate spamming contents which negatively impact on legitimate users. Spammers are motivated to launch various of attacks such as stealing personal information of users \citep{Bilge:09}, spreading viruses, malware \citep{grier2010spam}, phishing attacks, or compromise suspicious fake followers. Social spammers do not only annoy users, but also lead to financial loss and privacy issues of users. Therefore, the problem of social spamming is a serious issue prevalent on Twitter. Characterizing and detecting social spammers can keep Twitter as a spam-free environment and improve the quality of user experiences. \\\\
There are two main challenges of spammer detection on Twitter.
\begin{enumerate}
    \item The first challenge is how to process an enormous amount of Twitter data. Today, Twitter services handle more than 2.8 billion requests and store 4.5 petabytes of time series data every minutes \citep{twitter:observability}. We need an approach that can be able to scale up to handle a huge volume of data with limited computation capacity.
    \item Fast change of spamming patterns is the second challenge. The social spammer detection usually seems like a endless game between spammers and anti-spam systems. Spammers continually change their spamming strategy to fool the anti-spam systems. An approach that can be able to adapt to the complex, and fast changing of data is needed. 
\end{enumerate}
There were some previous studies on the Twitter spammer detection for years. Their approaches address this problem as the task of classifying a Twitter user into a spammer or a legitimate user. By analytic the spammers behaviors, they proposed effective features which related to content-based and network-based of users and then built a traditional batch learning model to detect spammers for future data. However, such batch learning models are less efficient due to rapid changing and quick evolution of spammers. Moreover, because of limited resources, it is very expensive if we gather all spamming patterns from batches to train the learning model. \\\\
One efficient approach for the fast evolve and large-scale of social spammer detection on Twitter is online learning. The online learning method continually updates the existing model while data arrives, as opposed to batch learning techniques which learn from the entire data.  In this paper, we study how to apply online learning on the social spammer detection on Twitter. Our experiment on Honey Pot dataset \citep{lee2011seven} and 1KS-10KN \citep{yang2012analyzing} dataset indicate the effectiveness of the online learning approach for this problem. \\\\
The main contributions of our work are outlined as follows:
\begin{enumerate}
    \item Successfully apply online learning for the problem of social spammer detection on Twitter. Our experiments show that online learning approaches efficiently reflect with the fast changing of data. 
    \item Find an optimal online learning method for social spammer detection on Twitter. We evaluated 16 online learning algorithms and find the Soft Confident-Weight algorithm achieves the best performance.  
    \item Evaluate the effectiveness of four different feature sets when applying online learning on this problem. The best result was observed by the combination of 2 sets of user network features and user activities features. This results indicated that user profile features and user content features are less robust than user network features and user activities features. 
\end{enumerate}
The remaining of this report is organized as follows. In the next section, we will present a brief overview of related works on spammer detection. A description of the methodology applied will be described in Section \ref{OL}. Then an experimental study showing the effectiveness of online learning is presented in Section \ref{exp}. Finally, we conclude and discuss future work in Section \ref{cons}.
\section{Related Works}
\subsection{Spammer detection in other platforms}
Spammers have been around us since the beginning of the electronic communication and adapted through the development of technology. Spam detection problem is a serious issue, and it has been studied for years on various platforms such as SMS \citep{GomezHidalgo:2006:CBS:1166160.1166191}, email \citep{Blanzieri:2008:SLT:1612711.1612715}, and the Web \citep{Webb06introducingthe}. A popular and well-developed approach for spam detection is based on machine learning techniques. They extracted effective features from historical data and built a supervised learning model. This model will be used to classify new data as either spam or legitimate user/message. 
\subsection{Spammer detection on Twitter}
As a result of the popularity of social media like Twitter, spammers are turning into the fast growing in this platform. There were some previous studies to tackle this problem. \\\\
Some studies focus on analyzing the spammer characteristics on Twitter. \citep{yardi2009detecting} explored the behavior of spammers from the entire life cycle of \#robotpickupline hashtag. According to their observations, spammers tend to send more messages and network interaction with others. Thus, the higher ratio of followers to followings of a user have, the higher probability that the user is a spammer. \citep{grier2010spam} studied perspective of spammers and click-through behaviors. Addition, they had already evaluated the effectiveness of backlists to prevent spamming. \citep{thomas2011suspended} collected the suspended accounts in 7 months from August 2010 to March 2011 and then they studied the characteristic of spammer account, tweet behavior, and spam campaign. \citep{Ghosh:12} focused on the link farm on Twitter by analyzing the suspended accounts. They observed that most of the link farms came from new users. They also proposed Collusion Rank to demote the ranking of link farms on Twitter. \\\\
\citep{Benevenuto10detectingspammers} addressed the study of spammer detection on Twitter trending topics. They used an SVM classifier to distinguish between spammers and legitimate users based on the basic of tweet contents and user profile information. Using the social honeypot to collected spammers is an interested work studied by \citep{Lee:10, lee2011seven}. After analysis spammer behavior, they extracted user profile features and user network features and built a supervised learning classifier to identify spammers. However, the approach requires a lot of time for observation spamming evidence. Addition, the collected data is often biased because it was only received content polluted from active spammers who was following the honeypot accounts. \\\\
\citep{yang2011free} observed that the proposed features from previous works were less effective with the evolving of spammers. They utilized ten new features for spammer detection on Twitter and evaluated these features with the existing ones. Their experiments indicated that using their new features give a better result for spammer detection problem. \citep{ferrara2014rise} used the information related to tweet content, user network, sentiment, and temporal patterns of activity for detecting Twitter bots. Some approaches have assessed the safety or suspiciousness of URLs in tweets as a mean to identify spam tweets \citep{thomas2011design, cao2015detecting, wang2013click, lee2012warningbird}. \\\\
More recent work has investigated the relationship between automation and spamming. In the study of \citep{amleshwaram2013cats}, a system for automated spammers detection is described. Features related to automation have been exploited to adapt to the changing structure of Twitters spammers population. An analysis of automated activity on Twitter was presented on \citep{chu2010tweeting}, and a system that detects the automation of an account is described in \citep{chu2012detectingIEEE}.\\\\
Previous works on spam detection on Twitter can be summarized as follows: collecting and analyzing spammer behaviors, define and extracting effective features, using supervised learning algorithms to build the statistical classifier to detect spammers. However, the behavior of spammers changes too fast in social media. It is hard for the batch-learning system to adapt to the evolving of spammers. 
\section{Online Learning for Social Spam Detection} \label{OL}
\begin{figure*}
\centering
\captionsetup{justification=centering}
\includegraphics[width=0.75\textwidth]{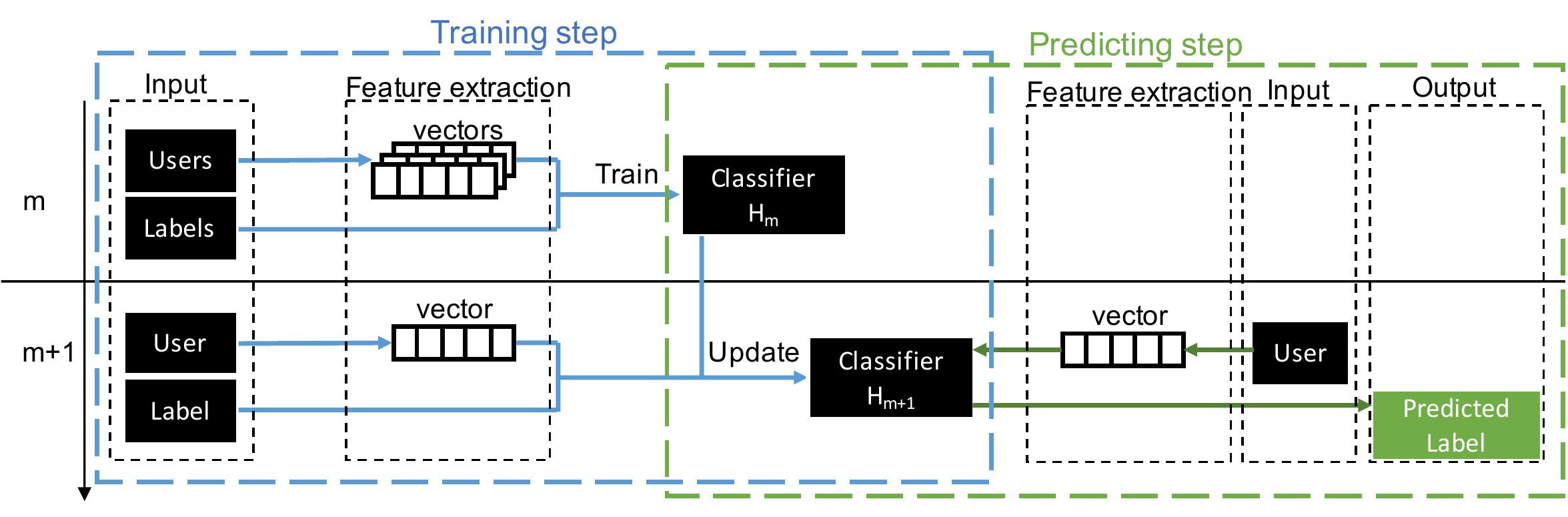}
\caption{The online learning framework for spammer detection on Twitter}
\label{fig:framework}
\end{figure*}

To detect spammer on Twitter, we propose a framework based on online supervised classification method. The classify model will be incrementally updated real-time. The overall framework is presented in the Fig \ref{fig:framework}. 
In the training step, given $m$ users with their identity label, the system will extract the features vector of each user. The detail of the extracted features will be talking in the feature represent spammers or legitimate users. Feature vectors and identity labels will be used to built the classifier model $H_{m}$. Given one more user, the feature extraction module will be used to extract the features of the user and update classifier model $H_{m}$ to $H_{m+1} $.
\subsection{Features represent spammers or legitimate users}
A spammer and a legitimate user have motivation differently in posting tweets or doing social activities. We can assume that the characteristics of spammers are quite different to legitimate users. The features to present a user include profile, social networks, activities and content of posted tweets. 
\subsubsection{User profile features}
The profile features are extracted from a user's Twitter profile and consist of:
\begin{itemize}
	\item the length of screen name
	\item whether the user profile has a description, the length of description
	\item whether the user profile has a URL 
	\item the longevity (age) of an account  (hours, days, weeks)
\end{itemize}
\subsubsection{User networks features}
The following features are used to characterize a user's social networks which mainly extracted from friendship information.
\begin{itemize}
	\item number of users following (friends) 
	\item number of followers 
	\item the ratio of number of following to number of followers
	\item the reputation of a user which is calculated by the ratio between the number of followers and the total number of followers and following. User's reputation $=  \frac{followers}{following + followers}$
	\item the following rate which is calculated the ratio between number of following and the longevity of account (hours, days, weeks)
	\item the followers rate which is calculated the ratio between number of followers and the longevity of account (hours, days, weeks)
	\item number of bidirectional friend $following \cap followes$
	\item the percentage of bidirectional friends with following $\frac{following \cap followes}{following}$
	\item the percentage of bidirectional friends with followers $\frac{following \cap followes}{followes}$
	\item the standard deviation of followings
	\item the standard deviation of followers
\end{itemize}
\subsubsection{User activity  features}
These features category capture user's social activities such as posted tweets or retweets. 
\begin{itemize}
	\item number of posted tweets
	\item number of posted tweets per hours, days, weeks
	\item number of content similarity of posted tweets by a user. 
	\item number of direct mentions (e.g., @username) per posted tweet
	\item number of direct mentions (e.g., @username) per hours, days, weeks
	\item number of URLs per tweet
	\item number of URLs per hours, days, weeks
	\item number of hashtags per tweet
	\item number of hashtags per hours, days, weeks
	\item number of retweets per tweet
	\item number of retweets per hours, days, weeks
\end{itemize}
\subsubsection{User content features}
Capture the linguistic properties of the text of each tweet such as part-of-speech tagging, the number of spam words from spamming words dictionary \footnote{https://github.com/splorp/wordpress-comment-blacklist}, Linguistic Inquiry and Word Count (LIWC) and sentiment features.
\begin{itemize}
    \item Part-of-speech tagging provides the syntactic information of a sentence and has been used in the natural language processing for measuring text informativeness. In detail implementation, we use the Twitter-specific tagger \citep{Gimpel:2011:PTT:2002736.2002747}. 
    \item Number of spam words is generated by matching a famous list of spam words. This list contains over 21,000 phrases, patterns, and keywords commonly used by spammers and comment bots in usernames, email addresses, link text. Since the masking behavior can dramatically decrease the proportion of spam tweets in a spamming account, applying this feature on an account content may not be helpful in detecting complex spamming accounts.
    \item LIWC dictionary is used to analyze text statistically and find psychologically-meaningful categories \citep{pennebaker2001linguistic}. There are 68 defined categories in LIWC dictionary. We use LIWC dictionary to compute 68 user's personality features. These features may help to determine the personality of spammers or legitimate users.
    \item In psychological, the micro expressions \citep{matsumoto2011evidence} play a distinct role in detecting deception. Inspired by this work, I explore the sentiment information could help capture deceptions of spammers. In this work, I use the list of lexicons from \citep{dodds2011temporal} for generating the sentiment features.
\end{itemize}
\subsection{Online learning algorithms}
This section briefly presents the online learning algorithms which we use for our evaluation. Informal, the online learning algorithms are trying to solve an online classification problem over a sequence of pairs $\{(x_1, y_1),$ $(x_2, y_{2}), ..., (x_{m}, y_{m})\}$, where each $x_{m}$ is an example's feature vector and $y_{m} \in \{0, 1\}$ is its label. At each step m during training, the algorithm makes a label prediction $h_{m}(x_{m})$, which for linear classifier is $h_{m}(x) = sign(w_{m}x)$.\\\\
After making a prediction, the algorithm receives the actual label $y_{m}$. Then the algorithms compute the loss $l(y_{m},\hat{y}_{m})$ based on some criterion to measure the difference between the prediction and the revealed true label $y_{m}$. The learner finally decides when and how to update the classification model at the end of each learning step bases on the result of the loss function.\\\\
In this paper, we have no vested interest in any particular strategy for online learning. We simply focus on the application of online learning on the problem social spammer detection on Twitter. \\\\
In our experiment, we use the LIBOL an online learning tool which was developed by \citep{hoi2014libol}. This tool consists of existing state-of-the-art online learning algorithms for large-scale online classification tasks. In details, these online learning algorithms can be grouped into two following categories: first-order online learning algorithms and second-order online learning algorithms. \\\\
First-order online learning algorithms include the algorithms that only keep updating one classification function. The examples algorithms in this categories are following:
\begin{itemize}
	\item Perceptron: the classical online learning algorithm \citep{rosenblatt1958perceptron}
	\item ALMA: the Approximate Maximal Margin algorithm \citep{gentile2002new}
	\item ROMMA: the Relaxed Online Maximum Margin algorithm \citep{li2002relaxed}
	\item OGD: the Online Gradient Descent algorithm \citep{zinkevich2003online}
	\item PA: the Passive Aggressive algorithm \citep{crammer2006online}
\end{itemize}
Second-order online learning algorithms have been explored in recent years. The major family of this categories assume the weight vector follows a Gaussian distribution $w \sim N(\mu, \Sigma)$ with mean vector $\mu \in \mathbb{R}^{d}$, covariance matrix $\Sigma \in \mathbb{R}^{d \times d}$ and dimensional vector space $d$. The examples algorithms in this categories are following:
\begin{itemize}
	\item SOP: the Second Order Perceptron algorithm \citep{cesa2005second}
	\item CW: the Confidence-Weight learning algorithm \citep{crammer2009exact}
	\item IELLIP: the online learning algorithms by improved ellipsoid method \citep{yang2009online}
	\item ARROW: the Adaptive Regularization of Weight Vectors algorithm \citep{crammer2009adaptive}
	\item NARROW: the New variant of Adaptive Regularization \citep{orabona2010new}
	\item NHERD: the Normal Herding method via Gaussian Herding \citep{crammer2010learning}
	\item SCW: the Soft Confidence Weight algorithms \citep{wang2012exact}
\end{itemize}
\section{Experiments} \label{exp}
In this section, we conduct the experiments to evaluate the effectiveness of online learning over the social spammer detection on Twitter. To demonstrate the effectiveness of online learning, we address the following questions:
\begin{enumerate}
	\item What is the accuracy of online learning methods compare with the batch learning methods on the single dataset?
	\item Do online learning algorithms provide a better adaptation on the data distribution changing?  
	\item Which online algorithms are most appropriate for our application? 
	\item How about the effectiveness of four different feature sets when applying online learning on the social spammer detection on Twitter?
\end{enumerate}
\subsection{Datasets}
In our experiments, we use two Twitter datasets. The statistics of two datasets are presented in Table \ref{table1-datasets}. 
\begin{table}
	\centering
	\caption{Statistics of 2 two datasets: Honey Pot \citep{lee2011seven} and 1KS-10KN \citep{yang2012analyzing}}
	\label{table1-datasets}
    \scalebox{0.75}{
	\begin{tabular}{l|r|r|r|r}
		\hline
		\multirow{2}{*}{}  & \multicolumn{2}{c|}{\textit{\textbf{Spammers}}}                                  & \multicolumn{2}{c}{\textit{\textbf{Legitimate Users}}}                         \\\\
		& \multicolumn{1}{l|}{\textbf{\# Users}} & \multicolumn{1}{l|}{\textbf{\# Tweets}} & \multicolumn{1}{l|}{\textbf{\# Users}} & \multicolumn{1}{l}{\textbf{\# Tweets}} \\\\ \hline
		\textbf{Honey Pot} & 22,223                                 & 2,353,473                               & 19,276                                 & 3,259,693                              \\\\
		\textbf{1KS-10KN}  & 1,000                                  & 145,095                                 & 10,000                                 & 1,209,521                              \\\\ \hline
	\end{tabular}}
\end{table}
\subsubsection{Social Honeypot Dataset:}  
\citep{lee2011seven} created 60 Tweeter accounts to attract spammers. After seven months, from December 2009 to August 2010, his team collected the information of 41,499 users and 5,613,166 their tweets. 22,223 users were labeled as spammers and 19,276 legitimate users. The ratio of spammers to legitimate users is around 1:1. 
\subsubsection{1KS-10KN Dataset:}
This dataset was collected by \citep{yang2012analyzing} from April 2010 to July 2010. This dataset contains the information of 11,000 users and 1,354,616 their tweets. They labeled 1,000 users as spammers and 10,000 legitimate users. The ratio of spammers to legitimate in the 1KS-10KN data is 1:10.\\\\
In two dataset, we don't have the information about the changing of network information of users through a specific time interval. We only have a snapshot of network information at the ending of collected data. For our work, we adapted the datasets with our purpose by randomly splitting each dataset to 20 parts and considering each part as a time interval.
\subsection{Effectiveness of Online learning on single dataset}
\begin{figure}
\centering
\subfigure[Honey Pot dataset]
{\label{fig:single-hp}\includegraphics[width=.75\linewidth]{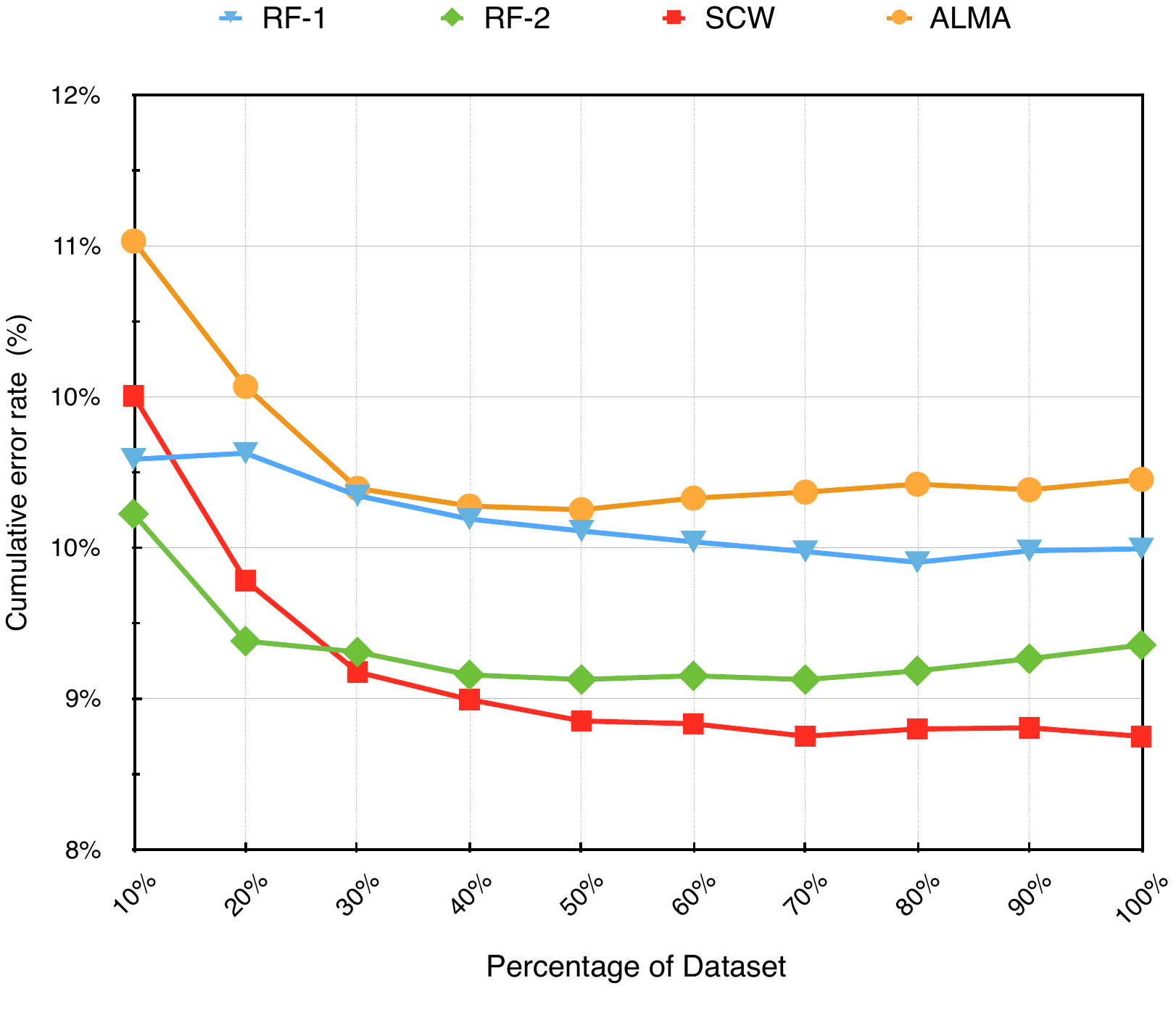}}
\subfigure[1KS-10KN dataset]
{\label{fig:single-1k}\includegraphics[width=.75\linewidth]{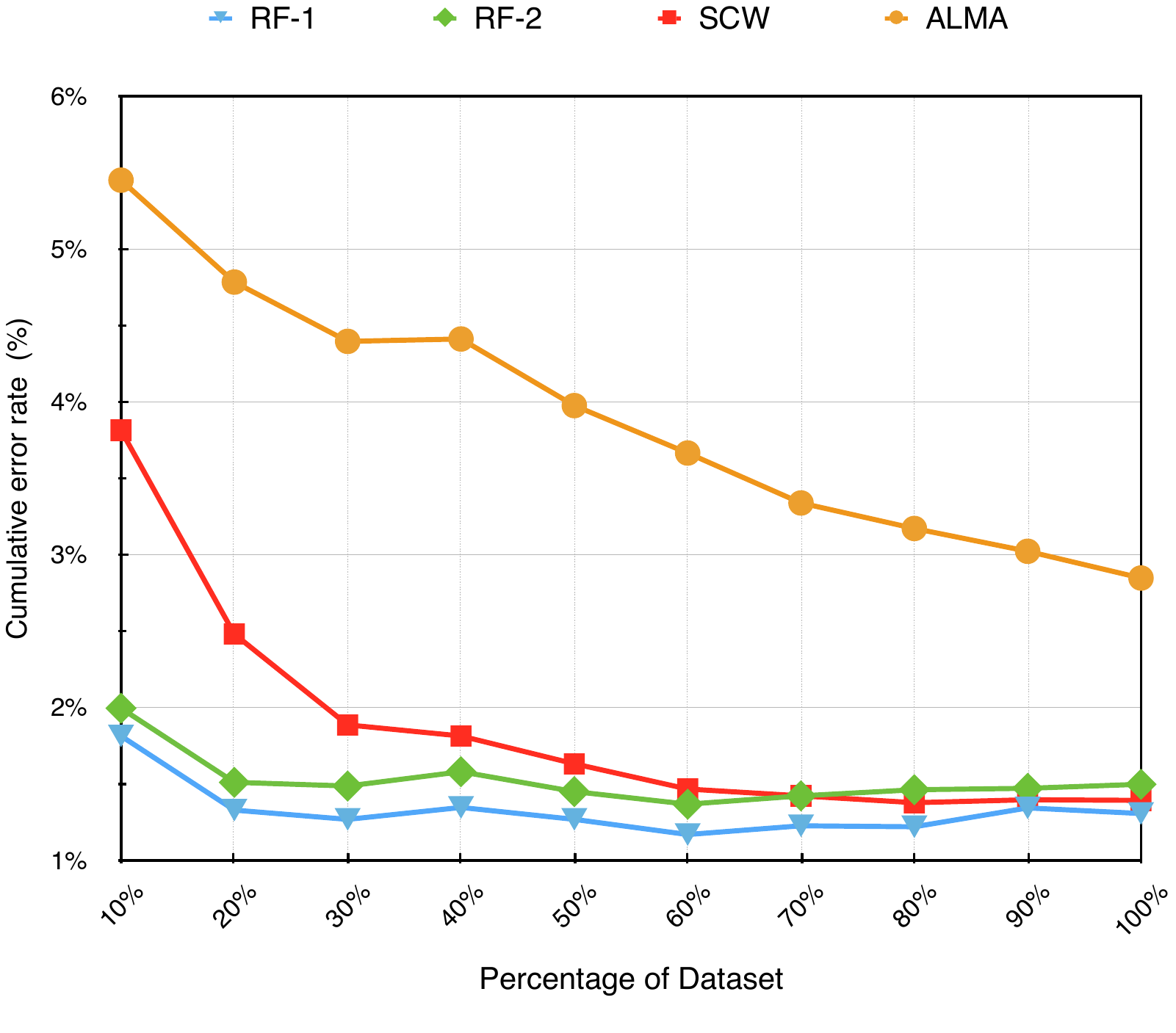}}
\caption{Effectiveness of online learning on single dataset}
\label{fig:single}
\end{figure}

In this section, we start by evaluating the effectiveness of online learning over the batch learning method regarding classification cumulative error rate on Honey Pot dataset and 1KS-10KN dataset. A lower the cumulative error rate means the better performance. In details, we compare two online learning algorithms - SCW \citep{wang2012exact} and ALMA \citep{gentile2002new} against two different training set configurations of the batch learning - Random Forest algorithms. We choose the batch learning - Random Forest algorithm because it was produced the highest performance on previous researches: \citep{lee2011seven, yang2011free}. 2 online learning algorithms: SCW \citep{wang2012exact} and ALMA \citep{gentile2002new} were chosen because they gave the lowest and second-lowest cumulative error rate in our experiments which will be presented in Table \ref{table:optimate-clf}. Assuming that our system can only process one part of the dataset, we conduct experiments to every single part of data. \\\\
Figure \ref{fig:single} and Table \ref{table:single} shows the classification cumulative error rates for online learning method and batch learning method on Honey Pot dataset and 1KS-10KN dataset. The x-axis shows the percentage of the dataset and the y-axis shows the cumulative error rate: percentage of miss-classified examples for all user up to this part. The RF-1 represent using the batch learning - Random Forest to train once on the first part of 20 splitting data. Then the model will be used to test on all the remaining parts. The RF-2 uses the same setting with RF-1 but retrain the model for every interval. For example, RF-2 train on part 2 of dataset and test on part 3. SCW and ALMA are 2 online learning used to make a single update over a cumulative training data. \\\\
In our experiment on single dataset, the distribution of data does not change much since we do not have the real streaming data on social spammer detection. We try to adapt the data Honey Pot and 1KS-10KN with our purpose. At the beginning part of data, the cumulative error rate of two online learning algorithms: SCW and ALMA are lower than batch learning: RF-1 and RF-2. However, the next results on the remaining parts show that SCW and ALMA faster reduce the cumulative error rate than batch learning. It means that online learning gives a more rapid adaptability when testing with other parts of data. \\\\
These results tested on single dataset indicate that online learning can give a comparable result with batch-learning when the distribution of data does not change much. Although online learning does not achieve the better result compare with batch learning techniques, it allows faster adapt to new data. \\\\
\begin{table}
	\centering
	\caption{Effectiveness of online learning on single dataset}
	\label{table:single}
    \scalebox{0.75}{
	\begin{tabular}{c|cccc|cccc}
		\hline
		\multirow{2}{*}{\%} & \multicolumn{4}{c|}{HoneyPot} & \multicolumn{4}{c}{1KS-10KN}  \\
		& RF1   & RF-2  & SCW   & ALMA  & RF-1  & RF-2  & SCW   & ALMA  \\ \hline
		10                  & 0.101 & 0.098 & 0.104 & 0.112 & 0.018 & 0.020 & 0.038 & 0.055 \\
		15                  & 0.102 & 0.095 & 0.099 & 0.108 & 0.014 & 0.018 & 0.030 & 0.053 \\
		20                  & 0.101 & 0.091 & 0.094 & 0.105 & 0.013 & 0.015 & 0.025 & 0.048 \\
		25                  & 0.099 & 0.090 & 0.092 & 0.102 & 0.012 & 0.014 & 0.021 & 0.044 \\
		30                  & 0.099 & 0.091 & 0.089 & 0.099 & 0.013 & 0.015 & 0.019 & 0.044 \\
		35                  & 0.099 & 0.090 & 0.090 & 0.099 & 0.014 & 0.015 & 0.019 & 0.044 \\
		40                  & 0.098 & 0.089 & 0.088 & 0.098 & 0.014 & 0.016 & 0.018 & 0.044 \\
		45                  & 0.098 & 0.090 & 0.088 & 0.099 & 0.013 & 0.015 & 0.017 & 0.042 \\
		50                  & 0.097 & 0.089 & 0.087 & 0.098 & 0.013 & 0.015 & 0.016 & 0.040 \\
		55                  & 0.097 & 0.089 & 0.087 & 0.099 & 0.012 & 0.014 & 0.016 & 0.039 \\
		60                  & 0.096 & 0.089 & 0.087 & 0.099 & 0.012 & 0.014 & 0.015 & 0.037 \\
		65                  & 0.097 & 0.090 & 0.087 & 0.099 & 0.012 & 0.014 & 0.014 & 0.035 \\
		70                  & 0.096 & 0.089 & 0.086 & 0.099 & 0.012 & 0.014 & 0.014 & 0.033 \\
		75                  & 0.095 & 0.089 & 0.086 & 0.099 & 0.012 & 0.015 & 0.014 & 0.032 \\
		80                  & 0.095 & 0.090 & 0.086 & 0.099 & 0.012 & 0.015 & 0.014 & 0.032 \\
		85                  & 0.096 & 0.090 & 0.086 & 0.099 & 0.013 & 0.015 & 0.014 & 0.031 \\
		90                  & 0.096 & 0.090 & 0.086 & 0.099 & 0.013 & 0.015 & 0.014 & 0.030 \\
		95                  & 0.096 & 0.090 & 0.086 & 0.099 & 0.013 & 0.015 & 0.014 & 0.029 \\
		100                 & 0.096 & 0.091 & 0.086 & 0.100 & 0.013 & 0.015 & 0.014 & 0.029 \\ \hline
	\end{tabular}}
\end{table}
\subsection{Effectiveness of Online learning on combine dataset}
\begin{figure}
\centering
\includegraphics[width=0.75\linewidth]{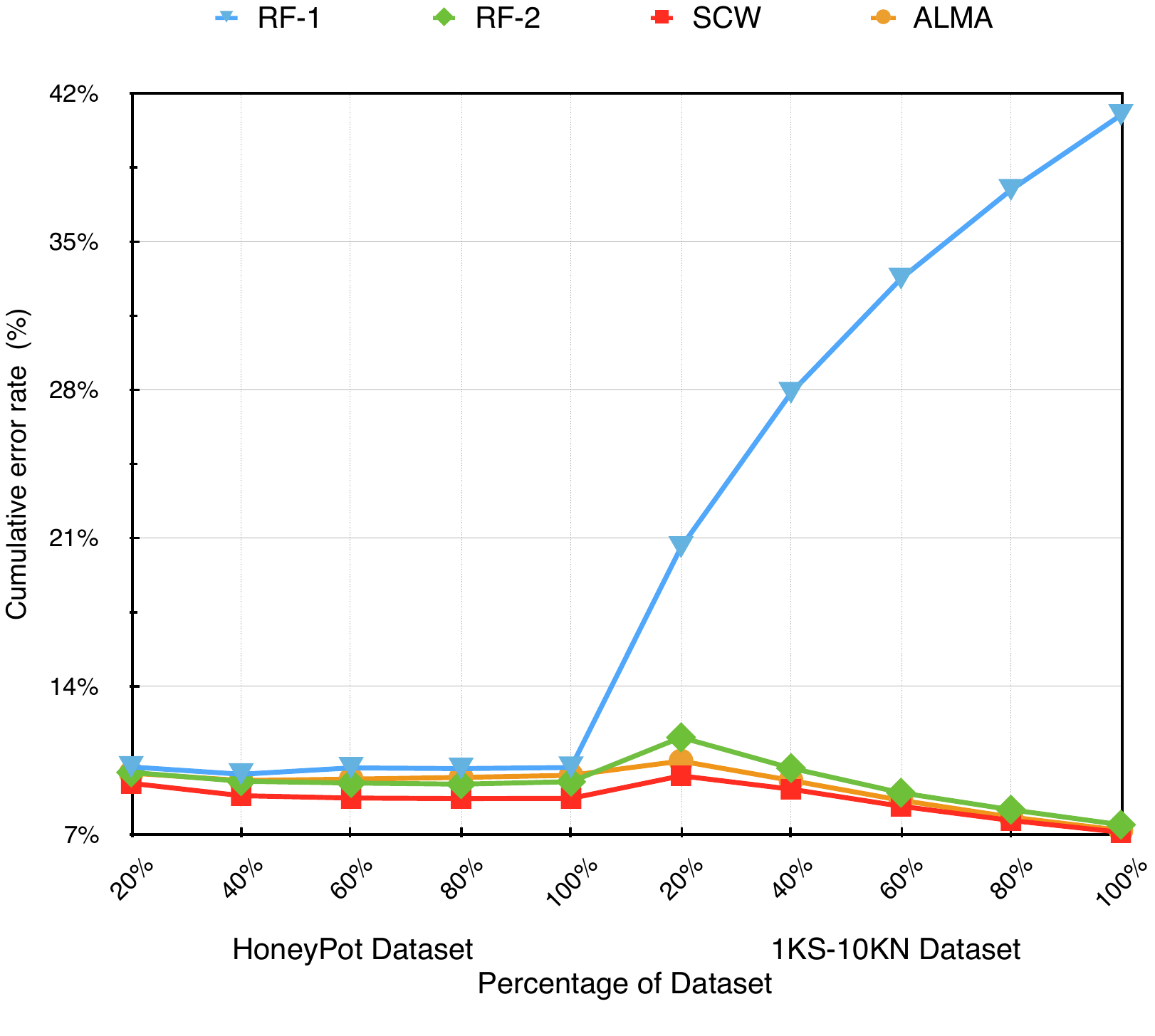}
\caption{Effectiveness of Online learning on combine dataset}
\label{fig:combine}
\end{figure}

In the real world problem, social spammers always change their spamming strategy; it makes rapid changing in the distribution of data. It explained why we study the ability to adapt to the changing distribution when to combine two datasets. We use the same online learning algorithms and two setting of batch learning in the previous section and evaluate on the combined dataset. In particular, this dataset contains 20 first parts from Honey Pot dataset and 20 next parts from 1KS-10KN dataset. \\\\
Figure \ref{fig:combine} show the result of social spammer detection on the combine of Honey Pot and 1KS-10KN dataset. In the RF-1, the cumulative error rate becomes significant increase when testing on data from 1KS-10KN. It is indicated the classifier model fail to detect spammers when data distribution changes. If we want to achieve better accuracy, we need to retrain the model with the new data from the 1KS-10KN dataset. In RF-2, the classifier will be re-trained from the prior interval, gave the better performance than RF-1. It means that the new data eventually can help to reduce the cumulative error rate. However, online learning algorithms - SCW and ALMA gave the better adaptability with the changing of data distribution since it gave the lowest cumulative error rate compare with the batch-learning method. The experiment results show that online learning very appropriate with the fast changing of spammers.
\begin{table}
	\centering
	\caption{Effectiveness of Online learning on combine dataset}
	\label{table:combine}
	\scalebox{0.75}{
		\begin{tabular}{@{}l|c|cccc@{}}
			\toprule
			       & \%  & RF-1  & RF-2  & SCW   & ALMA  \\ \midrule
			\multirow{19}{*}{\rotatebox[origin=c]{90}{Honey Pot}} & 10  & 0.102 & 0.101 & 0.105 & 0.107 \\
			& 15  & 0.103 & 0.102 & 0.098 & 0.102 \\
			& 20  & 0.102 & 0.100 & 0.094 & 0.099 \\
			& 25  & 0.102 & 0.096 & 0.093 & 0.098 \\
			& 30  & 0.100 & 0.096 & 0.091 & 0.096 \\
			& 35  & 0.101 & 0.097 & 0.091 & 0.097 \\
			& 40  & 0.099 & 0.095 & 0.089 & 0.096 \\
			& 45  & 0.101 & 0.097 & 0.089 & 0.096 \\
			& 50  & 0.102 & 0.096 & 0.087 & 0.096 \\
			& 55  & 0.102 & 0.095 & 0.087 & 0.096 \\
			& 60  & 0.102 & 0.095 & 0.087 & 0.096 \\
			& 65  & 0.102 & 0.095 & 0.087 & 0.097 \\
			& 70  & 0.102 & 0.094 & 0.087 & 0.097 \\
			& 75  & 0.102 & 0.094 & 0.087 & 0.097 \\
			& 80  & 0.101 & 0.094 & 0.087 & 0.097 \\
			& 85  & 0.101 & 0.095 & 0.087 & 0.097 \\
			& 90  & 0.101 & 0.095 & 0.087 & 0.098 \\
			& 95  & 0.101 & 0.095 & 0.087 & 0.098 \\
			& 100 & 0.102 & 0.095 & 0.087 & 0.098 \\
			\midrule
			\multirow{19}{*}{\rotatebox[origin=c]{90}{1KS - 10KN}} & 5   & 0.130 & 0.131 & 0.096 & 0.109 \\
			& 10  & 0.157 & 0.126 & 0.098 & 0.108 \\
			& 15  & 0.182 & 0.120 & 0.098 & 0.108 \\
			& 20  & 0.206 & 0.116 & 0.098 & 0.105 \\
			& 25  & 0.227 & 0.112 & 0.097 & 0.103 \\
			& 30  & 0.246 & 0.108 & 0.095 & 0.100 \\
			& 35  & 0.263 & 0.105 & 0.094 & 0.098 \\
			& 40  & 0.279 & 0.102 & 0.092 & 0.096 \\
			& 45  & 0.294 & 0.099 & 0.089 & 0.093 \\
			& 50  & 0.308 & 0.096 & 0.088 & 0.091 \\
			& 55  & 0.321 & 0.093 & 0.086 & 0.089 \\
			& 60  & 0.333 & 0.090 & 0.084 & 0.086 \\
			& 65  & 0.344 & 0.088 & 0.082 & 0.084 \\
			& 70  & 0.355 & 0.086 & 0.080 & 0.082 \\
			& 75  & 0.365 & 0.084 & 0.079 & 0.080 \\
			& 80  & 0.375 & 0.082 & 0.077 & 0.079 \\
			& 85  & 0.384 & 0.080 & 0.076 & 0.077 \\
			& 90  & 0.393 & 0.078 & 0.074 & 0.075 \\
			& 95  & 0.402 & 0.077 & 0.073 & 0.074 \\
			& 100 & 0.410 & 0.075 & 0.071 & 0.072 \\ \bottomrule
		\end{tabular}}
\end{table}
\subsection{Which online algorithms are most appropriate for social spammer detection}
In this section, we evaluate which of the online learning algorithms are the best suited to the social spammer detection. We tested 16 online learning algorithms implemented using the LIBOL tool. All of the experiments run on the combined dataset discussed in the previous section. \\\\
Table \ref{table:optimate-clf} shown the comparison of the performance of online learning algorithms. We find that the cumulative error rate ranged from 0.265 to 0.071. The Soft Confidence-Weight showed the best performance compare with the other online learning algorithms.
\begin{table*}
	\centering
	\caption{Performance comparison of online learning algorithms}
	\label{table:optimate-clf}
	\scalebox{0.75}{
	\begin{tabular}{c|c|cccccccccccccccc}
		\toprule
		\rotatebox[origin=l]{90}{}                      & \rotatebox[origin=l]{90}{\%}    & \rotatebox[origin=l]{90}{Perceptron} & \rotatebox[origin=l]{90}{ROMMA} & \rotatebox[origin=l]{90}{aROMMA} & \rotatebox[origin=l]{90}{ALMA}  & 
		\rotatebox[origin=l]{90}{OGD }  & 
		\rotatebox[origin=l]{90}{PA}    & 
		\rotatebox[origin=l]{90}{PA1}   & 
		\rotatebox[origin=l]{90}{PA2}   & 
		\rotatebox[origin=l]{90}{SOP}   & 
		\rotatebox[origin=l]{90}{IELLIP} & 
		\rotatebox[origin=l]{90}{CW}    &
		\rotatebox[origin=l]{90}{NHERD} & 
		\rotatebox[origin=l]{90}{AROW}  & 
		\rotatebox[origin=l]{90}{NAROW} & 
		\rotatebox[origin=l]{90}{SCW}   & 
		\rotatebox[origin=l]{90}{SCW2}  \\
		\midrule
		\multirow{19}{*}{\rotatebox[origin=c]{90}{Honey Pot}}  & 10\%  & 0.137      & 0.164 & 0.168  & 0.107 & 0.095 & 0.145 & 0.126 & 0.131 & 0.137 & 0.165  & 0.148 & 0.097 & 0.093 & 0.103 & 0.105 & 0.104 \\
		& 15\%  & 0.129      & 0.150 & 0.155  & 0.102 & 0.096 & 0.143 & 0.128 & 0.132 & 0.131 & 0.159  & 0.146 & 0.097 & 0.092 & 0.104 & 0.098 & 0.101 \\
		& 20\%  & 0.125      & 0.148 & 0.154  & 0.099 & 0.092 & 0.140 & 0.127 & 0.130 & 0.127 & 0.156  & 0.147 & 0.093 & 0.089 & 0.102 & 0.094 & 0.098 \\
		& 25\%  & 0.126      & 0.150 & 0.155  & 0.098 & 0.090 & 0.137 & 0.126 & 0.128 & 0.123 & 0.154  & 0.146 & 0.090 & 0.088 & 0.102 & 0.093 & 0.099 \\
		& 30\%  & 0.123      & 0.151 & 0.156  & 0.096 & 0.088 & 0.137 & 0.125 & 0.126 & 0.120 & 0.154  & 0.145 & 0.087 & 0.085 & 0.098 & 0.091 & 0.097 \\
		& 35\%  & 0.124      & 0.152 & 0.158  & 0.097 & 0.089 & 0.138 & 0.126 & 0.129 & 0.121 & 0.155  & 0.150 & 0.088 & 0.086 & 0.098 & 0.091 & 0.097 \\
		& 40\%  & 0.122      & 0.150 & 0.158  & 0.096 & 0.087 & 0.135 & 0.123 & 0.125 & 0.118 & 0.152  & 0.149 & 0.087 & 0.085 & 0.097 & 0.089 & 0.095 \\
		& 45\%  & 0.122      & 0.151 & 0.159  & 0.096 & 0.087 & 0.136 & 0.124 & 0.126 & 0.118 & 0.153  & 0.151 & 0.087 & 0.085 & 0.097 & 0.089 & 0.095 \\
		& 50\%  & 0.120      & 0.150 & 0.158  & 0.096 & 0.086 & 0.136 & 0.123 & 0.125 & 0.118 & 0.152  & 0.152 & 0.086 & 0.084 & 0.096 & 0.087 & 0.094 \\
		& 55\%  & 0.121      & 0.152 & 0.159  & 0.096 & 0.085 & 0.136 & 0.124 & 0.125 & 0.118 & 0.153  & 0.154 & 0.086 & 0.084 & 0.096 & 0.087 & 0.094 \\
		& 60\%  & 0.121      & 0.153 & 0.160  & 0.096 & 0.086 & 0.135 & 0.123 & 0.125 & 0.117 & 0.153  & 0.154 & 0.086 & 0.084 & 0.096 & 0.087 & 0.093 \\
		& 65\%  & 0.121      & 0.154 & 0.162  & 0.097 & 0.086 & 0.136 & 0.123 & 0.125 & 0.118 & 0.154  & 0.155 & 0.086 & 0.084 & 0.096 & 0.087 & 0.093 \\
		& 70\%  & 0.121      & 0.155 & 0.163  & 0.097 & 0.085 & 0.136 & 0.123 & 0.125 & 0.118 & 0.154  & 0.156 & 0.086 & 0.083 & 0.095 & 0.087 & 0.093 \\
		& 75\%  & 0.122      & 0.155 & 0.162  & 0.097 & 0.085 & 0.135 & 0.123 & 0.125 & 0.119 & 0.154  & 0.156 & 0.085 & 0.083 & 0.095 & 0.087 & 0.093 \\
		& 80\%  & 0.122      & 0.157 & 0.162  & 0.097 & 0.086 & 0.135 & 0.122 & 0.125 & 0.120 & 0.153  & 0.157 & 0.086 & 0.084 & 0.095 & 0.087 & 0.094 \\
		& 85\%  & 0.122      & 0.157 & 0.163  & 0.097 & 0.086 & 0.135 & 0.123 & 0.125 & 0.120 & 0.154  & 0.158 & 0.086 & 0.084 & 0.095 & 0.087 & 0.094 \\
		& 90\%  & 0.123      & 0.158 & 0.164  & 0.098 & 0.086 & 0.136 & 0.123 & 0.125 & 0.121 & 0.155  & 0.159 & 0.086 & 0.084 & 0.095 & 0.087 & 0.094 \\
		& 95\%  & 0.123      & 0.158 & 0.164  & 0.098 & 0.085 & 0.136 & 0.123 & 0.125 & 0.121 & 0.155  & 0.159 & 0.086 & 0.084 & 0.095 & 0.087 & 0.093 \\
		& 100\% & 0.123      & 0.159 & 0.165  & 0.098 & 0.086 & 0.136 & 0.123 & 0.126 & 0.121 & 0.155  & 0.161 & 0.086 & 0.085 & 0.095 & 0.087 & 0.094 \\
		\midrule
		\multirow{20}{*}{\rotatebox[origin=c]{90}{1KS - 10KN}} & 5\%   & 0.129      & 0.162 & 0.166  & 0.109 & 0.102 & 0.135 & 0.123 & 0.125 & 0.124 & 0.155  & 0.155 & 0.100 & 0.096 & 0.111 & 0.096 & 0.110 \\
		& 10\%  & 0.134      & 0.164 & 0.168  & 0.108 & 0.116 & 0.133 & 0.120 & 0.123 & 0.129 & 0.153  & 0.151 & 0.110 & 0.104 & 0.125 & 0.098 & 0.125 \\
		& 15\%  & 0.135      & 0.165 & 0.169  & 0.108 & 0.128 & 0.130 & 0.117 & 0.120 & 0.130 & 0.151  & 0.147 & 0.122 & 0.116 & 0.138 & 0.098 & 0.140 \\
		& 20\%  & 0.134      & 0.166 & 0.170  & 0.105 & 0.135 & 0.128 & 0.113 & 0.118 & 0.132 & 0.148  & 0.144 & 0.130 & 0.124 & 0.148 & 0.098 & 0.151 \\
		& 25\%  & 0.134      & 0.166 & 0.168  & 0.103 & 0.141 & 0.124 & 0.110 & 0.115 & 0.131 & 0.144  & 0.140 & 0.138 & 0.130 & 0.158 & 0.097 & 0.163 \\
		& 30\%  & 0.135      & 0.166 & 0.167  & 0.100 & 0.147 & 0.122 & 0.107 & 0.112 & 0.130 & 0.142  & 0.137 & 0.146 & 0.136 & 0.168 & 0.095 & 0.175 \\
		& 35\%  & 0.132      & 0.170 & 0.170  & 0.098 & 0.152 & 0.120 & 0.104 & 0.110 & 0.131 & 0.141  & 0.135 & 0.153 & 0.142 & 0.177 & 0.094 & 0.184 \\
		& 40\%  & 0.130      & 0.170 & 0.169  & 0.096 & 0.156 & 0.118 & 0.101 & 0.109 & 0.130 & 0.139  & 0.132 & 0.158 & 0.147 & 0.185 & 0.092 & 0.193 \\
		& 45\%  & 0.129      & 0.171 & 0.169  & 0.093 & 0.158 & 0.116 & 0.099 & 0.107 & 0.128 & 0.138  & 0.131 & 0.163 & 0.150 & 0.193 & 0.089 & 0.200 \\
		& 50\%  & 0.128      & 0.171 & 0.169  & 0.091 & 0.162 & 0.114 & 0.096 & 0.105 & 0.129 & 0.135  & 0.129 & 0.170 & 0.156 & 0.202 & 0.088 & 0.209 \\
		& 55\%  & 0.129      & 0.174 & 0.171  & 0.089 & 0.165 & 0.114 & 0.095 & 0.105 & 0.128 & 0.136  & 0.128 & 0.175 & 0.160 & 0.209 & 0.086 & 0.217 \\
		& 60\%  & 0.126      & 0.172 & 0.169  & 0.086 & 0.166 & 0.112 & 0.092 & 0.103 & 0.126 & 0.134  & 0.127 & 0.179 & 0.163 & 0.216 & 0.084 & 0.224 \\
		& 65\%  & 0.124      & 0.172 & 0.168  & 0.084 & 0.167 & 0.110 & 0.090 & 0.100 & 0.124 & 0.131  & 0.124 & 0.182 & 0.165 & 0.221 & 0.082 & 0.229 \\
		& 70\%  & 0.122      & 0.173 & 0.169  & 0.082 & 0.167 & 0.110 & 0.088 & 0.100 & 0.123 & 0.132  & 0.124 & 0.185 & 0.166 & 0.227 & 0.080 & 0.235 \\
		& 75\%  & 0.121      & 0.174 & 0.169  & 0.080 & 0.168 & 0.109 & 0.086 & 0.099 & 0.123 & 0.131  & 0.123 & 0.189 & 0.169 & 0.233 & 0.079 & 0.241 \\
		& 80\%  & 0.119      & 0.176 & 0.170  & 0.079 & 0.168 & 0.108 & 0.085 & 0.098 & 0.122 & 0.130  & 0.121 & 0.192 & 0.171 & 0.240 & 0.077 & 0.246 \\
		& 85\%  & 0.119      & 0.178 & 0.173  & 0.077 & 0.168 & 0.108 & 0.084 & 0.098 & 0.122 & 0.130  & 0.121 & 0.196 & 0.173 & 0.245 & 0.076 & 0.251 \\
		& 90\%  & 0.119      & 0.177 & 0.174  & 0.075 & 0.168 & 0.107 & 0.083 & 0.097 & 0.121 & 0.129  & 0.120 & 0.200 & 0.175 & 0.251 & 0.074 & 0.257 \\
		& 95\%  & 0.117      & 0.178 & 0.175  & 0.074 & 0.167 & 0.106 & 0.081 & 0.096 & 0.119 & 0.129  & 0.118 & 0.202 & 0.176 & 0.256 & 0.073 & 0.261 \\
		& 100\% & 0.115      & 0.180 & 0.176  & 0.072 & 0.166 & 0.107 & 0.080 & 0.097 & 0.118 & 0.129  & 0.118 & 0.204 & 0.176 & 0.262 & \textbf{0.071} & 0.265 \\
		\cmidrule(l){1-18}
	\end{tabular}}
\end{table*}
\subsection{Effectiveness of various feature set combinations}
\begin{figure}
\centering
\subfigure{\label{fig:cerscw}\includegraphics[width=.95\linewidth]{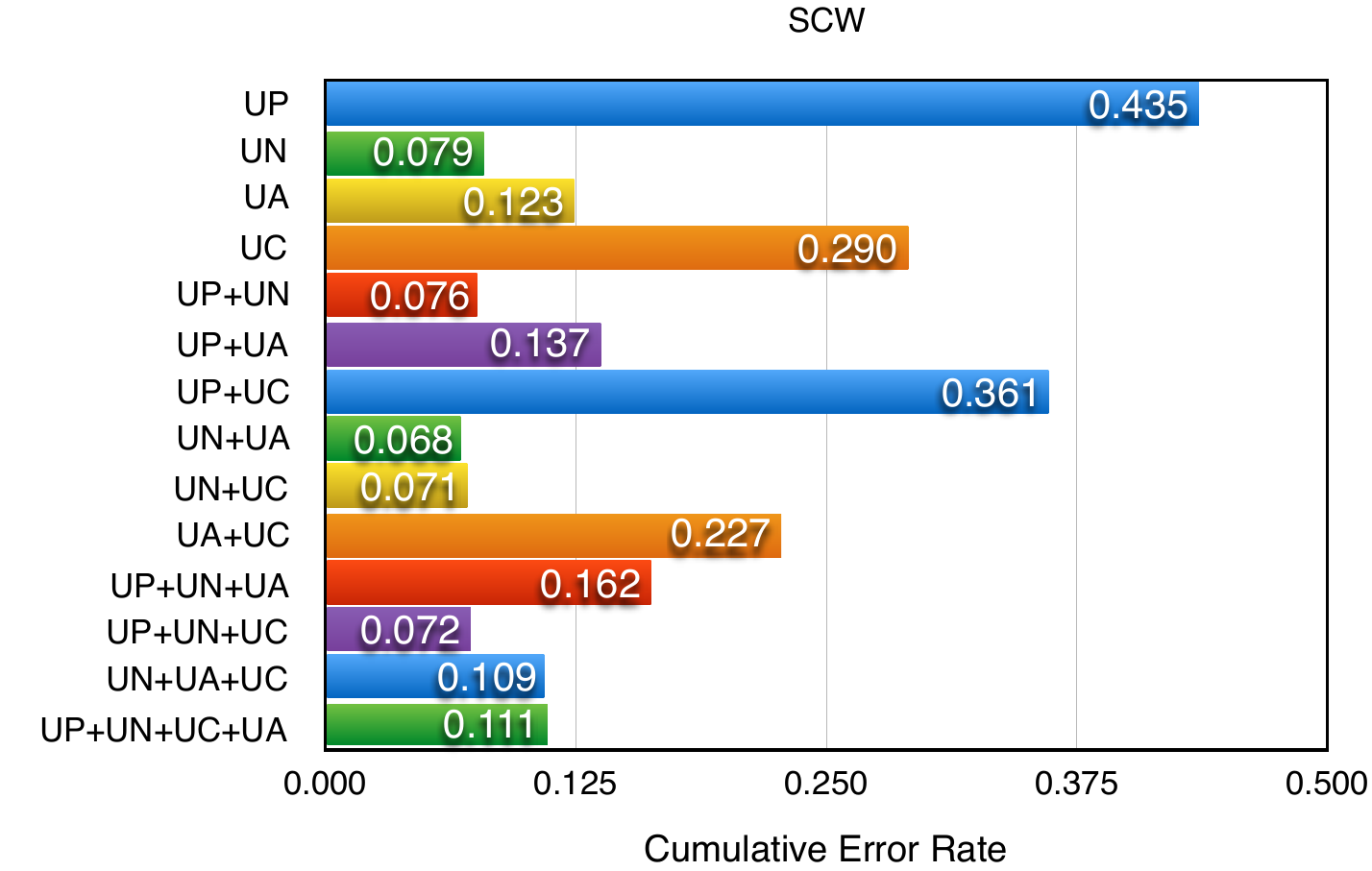}}
\subfigure{\label{fig:cerslma}\includegraphics[width=.95\linewidth]{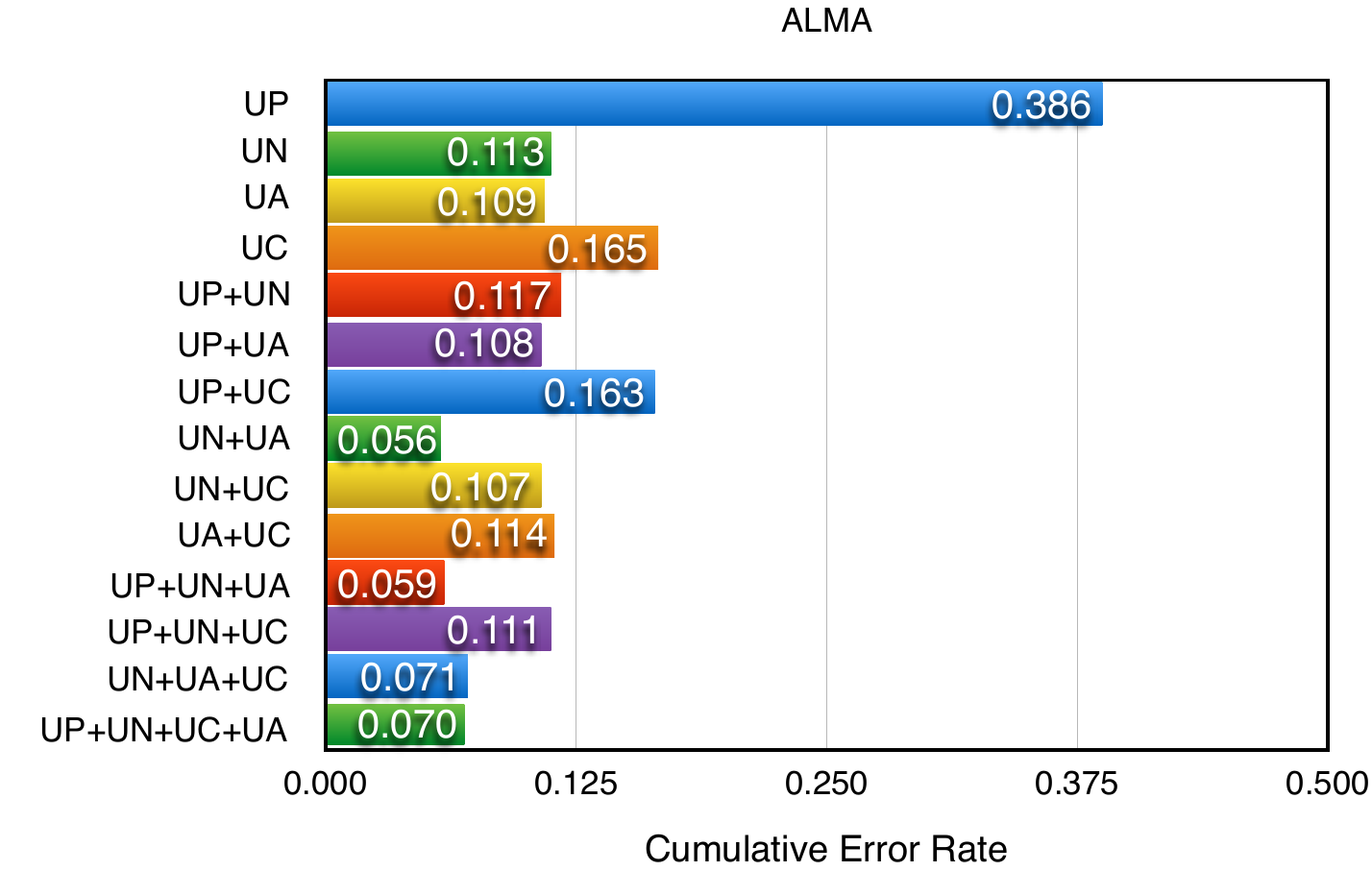}}
\caption{Evaluation of various feature set combination on SCW algorithm and ALMA algorithm}
\label{fig:cer}
\end{figure}

In this section, we trained 2 online learning (SCW-the highest performance and ALMA-the second-highest performance) with different feature sets (e.g. UP: User Profile features, UN: User Network features, UA: User Activities features, UC: User Content features) and various combination of the feature sets on the mixture of Honey Pot dataset and 1KS-10KN dataset. The experiment results are reported in Figure \ref{fig:cer}, Table \ref{table:featureSCW} and Table \ref{table:featureALMA}. \\\\
The result of social spammer detection by using SCW algorithm and ALMA algorithm are quite similar. The result of user profile features and user content features lower than user network features and user activity features when using a single feature set. It is consistent because social spammers are easy to fake their profile as a normal user on Twitter. Moreover, spammers quickly change their content information. It makes the user content features become noisy and less efficient. \\\\
In all of the experiments, the result of the combination of user network features and user activity features gives the best result in term of cumulative error rate 0.068 by using SCW algorithm and 0.056 by using ALMA. These results indicate that this two kind of features set are robustness with the changing spamming patterns and stable across the choice of online learning algorithms.
\begin{table*}[]
	\centering
	\caption{Evaluation of various feature set combinations on the SCW algorithm}
	\label{table:featureSCW}
	\scalebox{0.75}{
	\begin{tabular}{@{}c|c|cccccccccccccc@{}}
		\toprule
		\rotatebox[origin=l]{90}{}                     & 
		\rotatebox[origin=l]{90}{\%}    & 
		\rotatebox[origin=l]{90}{UP}    & 
		\rotatebox[origin=l]{90}{UN}    & 
		\rotatebox[origin=l]{90}{UA}    &
		\rotatebox[origin=l]{90}{UC}    &
		\rotatebox[origin=l]{90}{UP+UN} &
		\rotatebox[origin=l]{90}{UP+UA} &
		\rotatebox[origin=l]{90}{UP+UC} & 
		\rotatebox[origin=l]{90}{UN+UA} & 
		\rotatebox[origin=l]{90}{UN+UC} & 
		\rotatebox[origin=l]{90}{UA+UC} & 
		\rotatebox[origin=l]{90}{UP+UN+UA} & 
		\rotatebox[origin=l]{90}{UP+UN+UC} & 
		\rotatebox[origin=l]{90}{UN+UA+UC} & 
		\rotatebox[origin=l]{90}{UP+UN+UC+UA} \\ 
		\midrule
		\multirow{19}{*}{\rotatebox[origin=l]{90}{Honey Pot}} & 10\%  & 0.354 & 0.152 & 0.214 & 0.188 & 0.133 & 0.204 & 0.189 & 0.103 & 0.110 & 0.188 & 0.096    & 0.109    & 0.103    & 0.104       \\
		& 15\%  & 0.365 & 0.150 & 0.204 & 0.184 & 0.137 & 0.195 & 0.182 & 0.100 & 0.109 & 0.184 & 0.094    & 0.110    & 0.099    & 0.100       \\
		& 20\%  & 0.360 & 0.147 & 0.203 & 0.183 & 0.133 & 0.193 & 0.179 & 0.097 & 0.105 & 0.182 & 0.091    & 0.104    & 0.094    & 0.095       \\
		& 25\%  & 0.362 & 0.146 & 0.204 & 0.184 & 0.132 & 0.193 & 0.177 & 0.096 & 0.102 & 0.182 & 0.091    & 0.102    & 0.093    & 0.093       \\
		& 30\%  & 0.359 & 0.145 & 0.202 & 0.182 & 0.130 & 0.191 & 0.175 & 0.093 & 0.101 & 0.181 & 0.089    & 0.100    & 0.090    & 0.091       \\
		& 35\%  & 0.360 & 0.146 & 0.201 & 0.182 & 0.131 & 0.192 & 0.174 & 0.094 & 0.101 & 0.180 & 0.090    & 0.099    & 0.091    & 0.090       \\
		& 40\%  & 0.358 & 0.146 & 0.199 & 0.179 & 0.130 & 0.191 & 0.172 & 0.093 & 0.098 & 0.177 & 0.090    & 0.097    & 0.089    & 0.088       \\
		& 45\%  & 0.359 & 0.145 & 0.201 & 0.179 & 0.130 & 0.192 & 0.172 & 0.093 & 0.098 & 0.178 & 0.090    & 0.097    & 0.088    & 0.088       \\
		& 50\%  & 0.357 & 0.143 & 0.200 & 0.178 & 0.128 & 0.191 & 0.171 & 0.092 & 0.097 & 0.176 & 0.089    & 0.096    & 0.087    & 0.087       \\
		& 55\%  & 0.358 & 0.143 & 0.199 & 0.177 & 0.128 & 0.191 & 0.170 & 0.092 & 0.096 & 0.176 & 0.088    & 0.095    & 0.087    & 0.087       \\
		& 60\%  & 0.358 & 0.143 & 0.199 & 0.176 & 0.127 & 0.190 & 0.168 & 0.091 & 0.096 & 0.175 & 0.088    & 0.095    & 0.087    & 0.087       \\
		& 65\%  & 0.359 & 0.143 & 0.199 & 0.176 & 0.128 & 0.190 & 0.168 & 0.091 & 0.096 & 0.175 & 0.089    & 0.095    & 0.087    & 0.087       \\
		& 70\%  & 0.359 & 0.143 & 0.198 & 0.176 & 0.127 & 0.190 & 0.168 & 0.091 & 0.096 & 0.175 & 0.088    & 0.094    & 0.086    & 0.086       \\
		& 75\%  & 0.360 & 0.142 & 0.199 & 0.177 & 0.127 & 0.190 & 0.169 & 0.091 & 0.096 & 0.176 & 0.088    & 0.094    & 0.086    & 0.086       \\
		& 80\%  & 0.361 & 0.142 & 0.199 & 0.177 & 0.127 & 0.191 & 0.169 & 0.091 & 0.096 & 0.176 & 0.088    & 0.094    & 0.087    & 0.087       \\
		& 85\%  & 0.359 & 0.142 & 0.198 & 0.176 & 0.127 & 0.189 & 0.168 & 0.091 & 0.096 & 0.175 & 0.089    & 0.095    & 0.086    & 0.086       \\
		& 90\%  & 0.359 & 0.142 & 0.198 & 0.176 & 0.127 & 0.189 & 0.167 & 0.091 & 0.096 & 0.175 & 0.089    & 0.095    & 0.086    & 0.087       \\
		& 95\%  & 0.360 & 0.143 & 0.197 & 0.176 & 0.128 & 0.189 & 0.167 & 0.091 & 0.096 & 0.175 & 0.088    & 0.095    & 0.086    & 0.086       \\
		& 100\% & 0.360 & 0.143 & 0.198 & 0.175 & 0.128 & 0.189 & 0.168 & 0.091 & 0.096 & 0.175 & 0.089    & 0.095    & 0.086    & 0.086       \\
		\midrule
		\multirow{20}{*}{\rotatebox[origin=l]{90}{1KS-10KN}}  & 5\%   & 0.369 & 0.138 & 0.206 & 0.196 & 0.132 & 0.202 & 0.186 & 0.098 & 0.101 & 0.191 & 0.097    & 0.101    & 0.098    & 0.099       \\
		& 10\%  & 0.376 & 0.133 & 0.202 & 0.211 & 0.127 & 0.208 & 0.204 & 0.097 & 0.100 & 0.203 & 0.105    & 0.099    & 0.106    & 0.107       \\
		& 15\%  & 0.381 & 0.128 & 0.197 & 0.223 & 0.122 & 0.212 & 0.219 & 0.096 & 0.097 & 0.212 & 0.112    & 0.097    & 0.111    & 0.113       \\
		& 20\%  & 0.387 & 0.123 & 0.192 & 0.234 & 0.118 & 0.212 & 0.235 & 0.096 & 0.095 & 0.218 & 0.117    & 0.095    & 0.114    & 0.116       \\
		& 25\%  & 0.391 & 0.119 & 0.186 & 0.243 & 0.113 & 0.207 & 0.249 & 0.094 & 0.093 & 0.223 & 0.122    & 0.093    & 0.115    & 0.117       \\
		& 30\%  & 0.396 & 0.114 & 0.181 & 0.249 & 0.109 & 0.202 & 0.260 & 0.093 & 0.090 & 0.226 & 0.126    & 0.091    & 0.115    & 0.118       \\
		& 35\%  & 0.400 & 0.111 & 0.176 & 0.254 & 0.106 & 0.196 & 0.271 & 0.092 & 0.089 & 0.229 & 0.130    & 0.089    & 0.116    & 0.119       \\
		& 40\%  & 0.405 & 0.107 & 0.170 & 0.260 & 0.102 & 0.190 & 0.281 & 0.090 & 0.087 & 0.231 & 0.133    & 0.087    & 0.116    & 0.119       \\
		& 45\%  & 0.410 & 0.104 & 0.165 & 0.266 & 0.099 & 0.184 & 0.290 & 0.087 & 0.085 & 0.232 & 0.136    & 0.086    & 0.115    & 0.118       \\
		& 50\%  & 0.411 & 0.101 & 0.160 & 0.270 & 0.096 & 0.178 & 0.299 & 0.085 & 0.084 & 0.233 & 0.139    & 0.084    & 0.115    & 0.119       \\
		& 55\%  & 0.414 & 0.098 & 0.155 & 0.274 & 0.094 & 0.173 & 0.306 & 0.083 & 0.082 & 0.234 & 0.142    & 0.083    & 0.115    & 0.118       \\
		& 60\%  & 0.417 & 0.095 & 0.151 & 0.277 & 0.091 & 0.168 & 0.313 & 0.081 & 0.081 & 0.234 & 0.145    & 0.081    & 0.114    & 0.117       \\
		& 65\%  & 0.421 & 0.092 & 0.147 & 0.279 & 0.089 & 0.163 & 0.321 & 0.079 & 0.079 & 0.233 & 0.147    & 0.080    & 0.113    & 0.116       \\
		& 70\%  & 0.424 & 0.090 & 0.143 & 0.281 & 0.087 & 0.159 & 0.329 & 0.078 & 0.078 & 0.232 & 0.150    & 0.079    & 0.112    & 0.115       \\
		& 75\%  & 0.426 & 0.088 & 0.139 & 0.284 & 0.085 & 0.155 & 0.335 & 0.076 & 0.077 & 0.231 & 0.152    & 0.077    & 0.112    & 0.114       \\
		& 80\%  & 0.428 & 0.086 & 0.136 & 0.286 & 0.082 & 0.151 & 0.340 & 0.074 & 0.076 & 0.231 & 0.154    & 0.076    & 0.111    & 0.114       \\
		& 85\%  & 0.430 & 0.084 & 0.132 & 0.288 & 0.081 & 0.147 & 0.345 & 0.072 & 0.075 & 0.230 & 0.157    & 0.075    & 0.111    & 0.113       \\
		& 90\%  & 0.432 & 0.082 & 0.129 & 0.289 & 0.079 & 0.143 & 0.349 & 0.071 & 0.073 & 0.229 & 0.159    & 0.074    & 0.111    & 0.113       \\
		& 95\%  & 0.434 & 0.080 & 0.126 & 0.290 & 0.078 & 0.140 & 0.355 & 0.069 & 0.072 & 0.228 & 0.161    & 0.073    & 0.110    & 0.112       \\
		& 100\% & 0.435 & 0.079 & 0.123 & 0.290 & 0.076 & 0.137 & 0.361 & \textbf{0.068} & 0.071 & 0.227 & 0.162    & 0.072    & 0.109    & 0.111       \\ \cmidrule(l){1-16} 
	\end{tabular}}
\end{table*}

\begin{table*}
	\centering
	\caption{Evaluation of various feature set combinations on the ALMA algorithm}
	\label{table:featureALMA}
	\scalebox{0.75}{
	\begin{tabular}{@{}c|c|cccccccccccccc@{}}
			\toprule
			\rotatebox[origin=l]{90}{}                     & 
			\rotatebox[origin=l]{90}{\%}    & 
			\rotatebox[origin=l]{90}{UP}    & 
			\rotatebox[origin=l]{90}{UN}    & 
			\rotatebox[origin=l]{90}{UA}    &
			\rotatebox[origin=l]{90}{UC}    &
			\rotatebox[origin=l]{90}{UP+UN} &
			\rotatebox[origin=l]{90}{UP+UA} &
			\rotatebox[origin=l]{90}{UP+UC} & 
			\rotatebox[origin=l]{90}{UN+UA} & 
			\rotatebox[origin=l]{90}{UN+UC} & 
			\rotatebox[origin=l]{90}{UA+UC} & 
			\rotatebox[origin=l]{90}{UP+UN+UA} & 
			\rotatebox[origin=l]{90}{UP+UN+UC} & 
			\rotatebox[origin=l]{90}{UN+UA+UC} & 
			\rotatebox[origin=l]{90}{UP+UN+UC+UA} \\ 
			\midrule
		\multirow{19}{*}{\rotatebox[origin=l]{90}{Honey Pot}} & 10  & 0.359 & 0.137 & 0.227 & 0.205 & 0.132 & 0.207 & 0.201 & 0.101 & 0.126 & 0.197 & 0.096    & 0.122    & 0.103    & 0.107       \\
		& 15  & 0.362 & 0.140 & 0.216 & 0.203 & 0.135 & 0.196 & 0.194 & 0.103 & 0.123 & 0.193 & 0.096    & 0.121    & 0.101    & 0.105       \\
		& 20  & 0.360 & 0.135 & 0.213 & 0.202 & 0.131 & 0.196 & 0.192 & 0.100 & 0.120 & 0.194 & 0.095    & 0.117    & 0.098    & 0.102       \\
		& 25  & 0.360 & 0.134 & 0.213 & 0.200 & 0.131 & 0.196 & 0.190 & 0.100 & 0.118 & 0.193 & 0.095    & 0.114    & 0.097    & 0.100       \\
		& 30  & 0.359 & 0.131 & 0.213 & 0.198 & 0.128 & 0.194 & 0.186 & 0.097 & 0.115 & 0.192 & 0.091    & 0.111    & 0.094    & 0.096       \\
		& 35  & 0.360 & 0.132 & 0.212 & 0.197 & 0.129 & 0.196 & 0.185 & 0.099 & 0.113 & 0.191 & 0.093    & 0.111    & 0.095    & 0.096       \\
		& 40  & 0.359 & 0.132 & 0.210 & 0.193 & 0.127 & 0.194 & 0.180 & 0.098 & 0.111 & 0.188 & 0.091    & 0.108    & 0.093    & 0.095       \\
		& 45  & 0.360 & 0.132 & 0.211 & 0.196 & 0.127 & 0.197 & 0.183 & 0.098 & 0.111 & 0.189 & 0.093    & 0.108    & 0.094    & 0.095       \\
		& 50  & 0.357 & 0.130 & 0.211 & 0.194 & 0.126 & 0.196 & 0.182 & 0.097 & 0.110 & 0.188 & 0.092    & 0.107    & 0.094    & 0.095       \\
		& 55  & 0.359 & 0.131 & 0.211 & 0.194 & 0.126 & 0.197 & 0.183 & 0.097 & 0.109 & 0.188 & 0.091    & 0.107    & 0.094    & 0.095       \\
		& 60  & 0.359 & 0.130 & 0.210 & 0.193 & 0.126 & 0.196 & 0.181 & 0.096 & 0.109 & 0.187 & 0.092    & 0.107    & 0.094    & 0.095       \\
		& 65  & 0.360 & 0.130 & 0.211 & 0.193 & 0.126 & 0.196 & 0.182 & 0.096 & 0.109 & 0.187 & 0.092    & 0.108    & 0.094    & 0.095       \\
		& 70  & 0.360 & 0.130 & 0.210 & 0.193 & 0.125 & 0.196 & 0.182 & 0.096 & 0.108 & 0.187 & 0.092    & 0.108    & 0.094    & 0.094       \\
		& 75  & 0.359 & 0.129 & 0.211 & 0.193 & 0.125 & 0.196 & 0.182 & 0.096 & 0.109 & 0.188 & 0.092    & 0.108    & 0.094    & 0.095       \\
		& 80  & 0.359 & 0.130 & 0.212 & 0.193 & 0.125 & 0.196 & 0.182 & 0.097 & 0.109 & 0.187 & 0.092    & 0.108    & 0.095    & 0.095       \\
		& 85  & 0.357 & 0.130 & 0.211 & 0.192 & 0.125 & 0.195 & 0.180 & 0.097 & 0.108 & 0.187 & 0.093    & 0.108    & 0.095    & 0.095       \\
		& 90  & 0.356 & 0.130 & 0.210 & 0.192 & 0.125 & 0.195 & 0.180 & 0.097 & 0.109 & 0.187 & 0.092    & 0.109    & 0.095    & 0.095       \\
		& 95  & 0.357 & 0.130 & 0.210 & 0.193 & 0.125 & 0.195 & 0.180 & 0.098 & 0.109 & 0.187 & 0.093    & 0.109    & 0.095    & 0.096       \\
		& 100 & 0.357 & 0.131 & 0.210 & 0.193 & 0.125 & 0.195 & 0.181 & 0.098 & 0.110 & 0.187 & 0.093    & 0.109    & 0.095    & 0.096       \\
		\midrule
		\multirow{19}{*}{\rotatebox[origin=l]{90}{1KS-10KN}} & 5   & 0.364 & 0.144 & 0.203 & 0.198 & 0.139 & 0.198 & 0.191 & 0.100 & 0.120 & 0.189 & 0.102    & 0.120    & 0.106    & 0.105       \\
		& 10  & 0.370 & 0.152 & 0.194 & 0.198 & 0.149 & 0.191 & 0.193 & 0.096 & 0.122 & 0.186 & 0.100    & 0.124    & 0.107    & 0.105       \\
		& 15  & 0.375 & 0.158 & 0.185 & 0.197 & 0.156 & 0.183 & 0.193 & 0.092 & 0.122 & 0.181 & 0.097    & 0.125    & 0.106    & 0.103       \\
		& 20  & 0.379 & 0.160 & 0.178 & 0.196 & 0.160 & 0.175 & 0.192 & 0.088 & 0.122 & 0.176 & 0.095    & 0.126    & 0.104    & 0.101       \\
		& 25  & 0.381 & 0.161 & 0.171 & 0.194 & 0.162 & 0.168 & 0.190 & 0.085 & 0.121 & 0.171 & 0.093    & 0.125    & 0.102    & 0.099       \\
		& 30  & 0.383 & 0.159 & 0.164 & 0.191 & 0.161 & 0.162 & 0.188 & 0.082 & 0.120 & 0.166 & 0.091    & 0.125    & 0.099    & 0.097       \\
		& 35  & 0.385 & 0.157 & 0.158 & 0.189 & 0.160 & 0.157 & 0.186 & 0.079 & 0.120 & 0.161 & 0.086    & 0.124    & 0.097    & 0.094       \\
		& 40  & 0.385 & 0.153 & 0.153 & 0.186 & 0.157 & 0.151 & 0.184 & 0.077 & 0.118 & 0.156 & 0.083    & 0.123    & 0.095    & 0.092       \\
		& 45  & 0.387 & 0.148 & 0.148 & 0.184 & 0.152 & 0.146 & 0.182 & 0.074 & 0.117 & 0.152 & 0.080    & 0.122    & 0.092    & 0.090       \\
		& 50  & 0.389 & 0.143 & 0.143 & 0.182 & 0.148 & 0.141 & 0.180 & 0.072 & 0.116 & 0.147 & 0.079    & 0.121    & 0.090    & 0.088       \\
		& 55  & 0.391 & 0.139 & 0.138 & 0.181 & 0.144 & 0.137 & 0.179 & 0.070 & 0.115 & 0.143 & 0.077    & 0.120    & 0.088    & 0.085       \\
		& 60  & 0.391 & 0.135 & 0.134 & 0.178 & 0.140 & 0.133 & 0.177 & 0.068 & 0.113 & 0.139 & 0.075    & 0.119    & 0.085    & 0.083       \\
		& 65  & 0.391 & 0.132 & 0.130 & 0.176 & 0.137 & 0.129 & 0.175 & 0.066 & 0.113 & 0.135 & 0.073    & 0.118    & 0.083    & 0.081       \\
		& 70  & 0.390 & 0.129 & 0.127 & 0.174 & 0.134 & 0.125 & 0.173 & 0.064 & 0.112 & 0.132 & 0.071    & 0.117    & 0.081    & 0.079       \\
		& 75  & 0.390 & 0.126 & 0.123 & 0.172 & 0.131 & 0.122 & 0.171 & 0.063 & 0.111 & 0.129 & 0.067    & 0.116    & 0.079    & 0.077       \\
		& 80  & 0.390 & 0.123 & 0.120 & 0.170 & 0.128 & 0.119 & 0.169 & 0.061 & 0.110 & 0.126 & 0.065    & 0.115    & 0.077    & 0.076       \\
		& 85  & 0.390 & 0.120 & 0.117 & 0.169 & 0.125 & 0.116 & 0.168 & 0.060 & 0.110 & 0.123 & 0.063    & 0.114    & 0.076    & 0.074       \\
		& 90  & 0.389 & 0.118 & 0.114 & 0.168 & 0.123 & 0.113 & 0.166 & 0.059 & 0.109 & 0.120 & 0.062    & 0.113    & 0.074    & 0.073       \\
		& 95  & 0.389 & 0.115 & 0.112 & 0.166 & 0.120 & 0.110 & 0.165 & 0.058 & 0.108 & 0.117 & 0.060    & 0.112    & 0.073    & 0.071       \\
		& 100 & 0.386 & 0.113 & 0.109 & 0.165 & 0.117 & 0.108 & 0.163 & \textbf{0.056} & 0.107 & 0.114 & 0.059    & 0.111    & 0.071    & 0.070       \\ \bottomrule
	\end{tabular}}
\end{table*}
\section{Conclusion and Future Work} \label{cons}
The social spammer detection on Twitter is sophisticated and adaptable to game the system by continually change their content and network patterns. To handle fast evolving social spammers, we suggest using the online learning incrementally update classifier model when the newly spamming pattern occurred. Our experiment results show that the approach is effective in dynamically changing spamming strategies of spammers comparing with other batch learning method. Additionally, we address that the online learning - Soft Confident-Weight achieve the best result compare with other online algorithms. We also studied the effectiveness of four feature sets on two online learning algorithms - SCW and ALMA. The experiments show that user network features and user activities features are more robustness than user profile features and user content features and stable across online learning algorithms.\\\\
In near future, the amount of available data has risen steadily. It imposes a computational burden on the single system. In consequences, we need an approach can be able to perform in a distributed fashion. With this motivation, the future works will be focused on how to build the scalable distributed spammer detection system.

\bibliographystyle{spbasic}      
\bibliography{ref}   

\begin{thebibliography}{40}
\providecommand{\natexlab}[1]{#1}
\providecommand{\url}[1]{{#1}}
\providecommand{\urlprefix}{URL }
\expandafter\ifx\csname urlstyle\endcsname\relax
  \providecommand{\doi}[1]{DOI~\discretionary{}{}{}#1}\else
  \providecommand{\doi}{DOI~\discretionary{}{}{}\begingroup
  \urlstyle{rm}\Url}\fi
\providecommand{\eprint}[2][]{\url{#2}}

\bibitem[{Amleshwaram et~al(2013)Amleshwaram, Reddy, Yadav, Gu, and
  Yang}]{amleshwaram2013cats}
Amleshwaram AA, Reddy N, Yadav S, Gu G, Yang C (2013) Cats: Characterizing
  automation of twitter spammers. In: Communication Systems and Networks
  (COMSNETS), 2013 Fifth International Conference on, IEEE, pp 1--10

\bibitem[{Benevenuto et~al(2010)Benevenuto, Magno, Rodrigues, and
  Almeida}]{Benevenuto10detectingspammers}
Benevenuto F, Magno G, Rodrigues T, Almeida V (2010) Detecting spammers on
  twitter. In: In Collaboration, Electronic messaging, Anti-Abuse and Spam
  Conference (CEAS

\bibitem[{Bilge et~al(2009)Bilge, Strufe, Balzarotti, and Kirda}]{Bilge:09}
Bilge L, Strufe T, Balzarotti D, Kirda E (2009) All your contacts are belong to
  us: Automated identity theft attacks on social networks. In: Proceedings of
  the 18th International Conference on World Wide Web, ACM, New York, NY, USA,
  WWW '09, pp 551--560, \doi{10.1145/1526709.1526784},
  \urlprefix\url{http://doi.acm.org/10.1145/1526709.1526784}

\bibitem[{Blanzieri and Bryl(2008)}]{Blanzieri:2008:SLT:1612711.1612715}
Blanzieri E, Bryl A (2008) A survey of learning-based techniques of email spam
  filtering. Artif Intell Rev 29(1):63--92, \doi{10.1007/s10462-009-9109-6},
  \urlprefix\url{http://dx.doi.org/10.1007/s10462-009-9109-6}

\bibitem[{Cao and Caverlee(2015)}]{cao2015detecting}
Cao C, Caverlee J (2015) Detecting spam urls in social media via behavioral
  analysis. In: Advances in Information Retrieval, Springer, pp 703--714

\bibitem[{Cesa-Bianchi et~al(2005)Cesa-Bianchi, Conconi, and
  Gentile}]{cesa2005second}
Cesa-Bianchi N, Conconi A, Gentile C (2005) A second-order perceptron
  algorithm. SIAM Journal on Computing 34(3):640--668

\bibitem[{Chu et~al(2010)Chu, Gianvecchio, Wang, and Jajodia}]{chu2010tweeting}
Chu Z, Gianvecchio S, Wang H, Jajodia S (2010) Who is tweeting on twitter:
  human, bot, or cyborg? In: Proceedings of the 26th annual computer security
  applications conference, ACM, pp 21--30

\bibitem[{Chu et~al(2012)Chu, Gianvecchio, Wang, and
  Jajodia}]{chu2012detectingIEEE}
Chu Z, Gianvecchio S, Wang H, Jajodia S (2012) Detecting automation of twitter
  accounts: Are you a human, bot, or cyborg? Dependable and Secure Computing,
  IEEE Transactions on 9(6):811--824

\bibitem[{Crammer and Lee(2010)}]{crammer2010learning}
Crammer K, Lee DD (2010) Learning via gaussian herding. In: Advances in neural
  information processing systems, pp 451--459

\bibitem[{Crammer et~al(2006)Crammer, Dekel, Keshet, Shalev-Shwartz, and
  Singer}]{crammer2006online}
Crammer K, Dekel O, Keshet J, Shalev-Shwartz S, Singer Y (2006) Online
  passive-aggressive algorithms. The Journal of Machine Learning Research
  7:551--585

\bibitem[{Crammer et~al(2009{\natexlab{a}})Crammer, Dredze, and
  Pereira}]{crammer2009exact}
Crammer K, Dredze M, Pereira F (2009{\natexlab{a}}) Exact convex
  confidence-weighted learning. In: Advances in Neural Information Processing
  Systems, pp 345--352

\bibitem[{Crammer et~al(2009{\natexlab{b}})Crammer, Kulesza, and
  Dredze}]{crammer2009adaptive}
Crammer K, Kulesza A, Dredze M (2009{\natexlab{b}}) Adaptive regularization of
  weight vectors. In: Advances in neural information processing systems, pp
  414--422

\bibitem[{Dodds et~al(2011)Dodds, Harris, Kloumann, Bliss, and
  Danforth}]{dodds2011temporal}
Dodds PS, Harris KD, Kloumann IM, Bliss CA, Danforth CM (2011) Temporal
  patterns of happiness and information in a global social network:
  Hedonometrics and twitter. PloS one 6(12):e26,752

\bibitem[{Ferrara et~al(2014)Ferrara, Varol, Davis, Menczer, and
  Flammini}]{ferrara2014rise}
Ferrara E, Varol O, Davis C, Menczer F, Flammini A (2014) The rise of social
  bots. arXiv preprint arXiv:14075225

\bibitem[{Gentile(2002)}]{gentile2002new}
Gentile C (2002) A new approximate maximal margin classification algorithm. The
  Journal of Machine Learning Research 2:213--242

\bibitem[{Ghosh et~al(2012)Ghosh, Viswanath, Kooti, Sharma, Korlam, Benevenuto,
  Ganguly, and Gummadi}]{Ghosh:12}
Ghosh S, Viswanath B, Kooti F, Sharma NK, Korlam G, Benevenuto F, Ganguly N,
  Gummadi KP (2012) Understanding and combating link farming in the twitter
  social network. In: Proceedings of the 21st International Conference on World
  Wide Web, ACM, New York, NY, USA, WWW '12, pp 61--70,
  \doi{10.1145/2187836.2187846},
  \urlprefix\url{http://doi.acm.org/10.1145/2187836.2187846}

\bibitem[{Gimpel et~al(2011)Gimpel, Schneider, O'Connor, Das, Mills,
  Eisenstein, Heilman, Yogatama, Flanigan, and
  Smith}]{Gimpel:2011:PTT:2002736.2002747}
Gimpel K, Schneider N, O'Connor B, Das D, Mills D, Eisenstein J, Heilman M,
  Yogatama D, Flanigan J, Smith NA (2011) Part-of-speech tagging for twitter:
  Annotation, features, and experiments. In: Proceedings of the 49th Annual
  Meeting of the Association for Computational Linguistics: Human Language
  Technologies: Short Papers - Volume 2, Association for Computational
  Linguistics, Stroudsburg, PA, USA, HLT '11, pp 42--47,
  \urlprefix\url{http://dl.acm.org/citation.cfm?id=2002736.2002747}

\bibitem[{G\'{o}mez~Hidalgo et~al(2006)G\'{o}mez~Hidalgo, Bringas, S\'{a}nz,
  and Garc\'{\i}a}]{GomezHidalgo:2006:CBS:1166160.1166191}
G\'{o}mez~Hidalgo JM, Bringas GC, S\'{a}nz EP, Garc\'{\i}a FC (2006) Content
  based sms spam filtering. In: Proceedings of the 2006 ACM Symposium on
  Document Engineering, ACM, New York, NY, USA, DocEng '06, pp 107--114,
  \doi{10.1145/1166160.1166191},
  \urlprefix\url{http://doi.acm.org/10.1145/1166160.1166191}

\bibitem[{Grier et~al(2010)Grier, Thomas, Paxson, and Zhang}]{grier2010spam}
Grier C, Thomas K, Paxson V, Zhang M (2010) @ spam: the underground on 140
  characters or less. In: Proceedings of the 17th ACM conference on Computer
  and communications security, ACM, pp 27--37

\bibitem[{Hoi et~al(2014)Hoi, Wang, and Zhao}]{hoi2014libol}
Hoi SC, Wang J, Zhao P (2014) Libol: A library for online learning algorithms.
  The Journal of Machine Learning Research 15:495--499,
  \urlprefix\url{http://LIBOL.stevenhoi.org}

\bibitem[{Lee et~al(2010)Lee, Caverlee, and Webb}]{Lee:10}
Lee K, Caverlee J, Webb S (2010) Uncovering social spammers: Social honeypots +
  machine learning. In: Proceedings of the 33rd International ACM SIGIR
  Conference on Research and Development in Information Retrieval, ACM, New
  York, NY, USA, SIGIR '10, pp 435--442, \doi{10.1145/1835449.1835522},
  \urlprefix\url{http://doi.acm.org/10.1145/1835449.1835522}

\bibitem[{Lee et~al(2011)Lee, Eoff, and Caverlee}]{lee2011seven}
Lee K, Eoff BD, Caverlee J (2011) Seven months with the devils: A long-term
  study of content polluters on twitter. In: ICWSM, Citeseer

\bibitem[{Lee and Kim(2012)}]{lee2012warningbird}
Lee S, Kim J (2012) Warningbird: Detecting suspicious urls in twitter stream.
  In: NDSS

\bibitem[{Li and Long(2002)}]{li2002relaxed}
Li Y, Long PM (2002) The relaxed online maximum margin algorithm. Machine
  Learning 46(1-3):361--387

\bibitem[{Matsumoto and Hwang(2011)}]{matsumoto2011evidence}
Matsumoto D, Hwang HS (2011) Evidence for training the ability to read
  microexpressions of emotion. Motivation and Emotion 35(2):181--191

\bibitem[{Orabona and Crammer(2010)}]{orabona2010new}
Orabona F, Crammer K (2010) New adaptive algorithms for online classification.
  In: Advances in neural information processing systems, pp 1840--1848

\bibitem[{Pennebaker et~al(2001)Pennebaker, Francis, and
  Booth}]{pennebaker2001linguistic}
Pennebaker JW, Francis ME, Booth RJ (2001) Linguistic inquiry and word count:
  Liwc 2001. Mahway: Lawrence Erlbaum Associates 71:2001

\bibitem[{Rosenblatt(1958)}]{rosenblatt1958perceptron}
Rosenblatt F (1958) The perceptron: a probabilistic model for information
  storage and organization in the brain. Psychological review 65(6):386

\bibitem[{Thomas et~al(2011{\natexlab{a}})Thomas, Grier, Ma, Paxson, and
  Song}]{thomas2011design}
Thomas K, Grier C, Ma J, Paxson V, Song D (2011{\natexlab{a}}) Design and
  evaluation of a real-time url spam filtering service. In: Security and
  Privacy (SP), 2011 IEEE Symposium on, IEEE, pp 447--462

\bibitem[{Thomas et~al(2011{\natexlab{b}})Thomas, Grier, Song, and
  Paxson}]{thomas2011suspended}
Thomas K, Grier C, Song D, Paxson V (2011{\natexlab{b}}) Suspended accounts in
  retrospect: an analysis of twitter spam. In: Proceedings of the 2011 ACM
  SIGCOMM conference on Internet measurement conference, ACM, pp 243--258

\bibitem[{Twitter(2016)}]{twitter:observability}
Twitter (2016) Observability at twitter: technical overview, part i

\bibitem[{Wang et~al(2013)Wang, Navathe, Liu, Irani, Tamersoy, and
  Pu}]{wang2013click}
Wang D, Navathe SB, Liu L, Irani D, Tamersoy A, Pu C (2013) Click traffic
  analysis of short url spam on twitter. In: Collaborative Computing:
  Networking, Applications and Worksharing (Collaboratecom), 2013 9th
  International Conference Conference on, IEEE, pp 250--259

\bibitem[{Wang et~al(2012)Wang, Zhao, and Hoi}]{wang2012exact}
Wang J, Zhao P, Hoi SC (2012) Exact soft confidence-weighted learning. arXiv
  preprint arXiv:12064612

\bibitem[{Webb(2006)}]{Webb06introducingthe}
Webb S (2006) Introducing the webb spam corpus: Using email spam to identify
  web spam automatically. In: In Proceedings of the 3rd Conference on Email and
  AntiSpam (CEAS) (Mountain View

\bibitem[{Wikipedia(2016)}]{wiki:spamming}
Wikipedia (2016) Spamming --- wikipedia{,} the free encyclopedia.
  \urlprefix\url{https://en.wikipedia.org/w/index.php?title=Spamming&oldid=710141595},
  [Online; accessed 25-March-2016]

\bibitem[{Yang et~al(2011)Yang, Harkreader, and Gu}]{yang2011free}
Yang C, Harkreader RC, Gu G (2011) Die free or live hard? empirical evaluation
  and new design for fighting evolving twitter spammers. In: Recent Advances in
  Intrusion Detection, Springer, pp 318--337

\bibitem[{Yang et~al(2012)Yang, Harkreader, Zhang, Shin, and
  Gu}]{yang2012analyzing}
Yang C, Harkreader R, Zhang J, Shin S, Gu G (2012) Analyzing spammers' social
  networks for fun and profit: a case study of cyber criminal ecosystem on
  twitter. In: Proceedings of the 21st international conference on World Wide
  Web, ACM, pp 71--80

\bibitem[{Yang et~al(2009)Yang, Jin, and Ye}]{yang2009online}
Yang L, Jin R, Ye J (2009) Online learning by ellipsoid method. In: Proceedings
  of the 26th Annual International Conference on Machine Learning, ACM, pp
  1153--1160

\bibitem[{Yardi et~al(2009)Yardi, Romero, Schoenebeck
  et~al}]{yardi2009detecting}
Yardi S, Romero D, Schoenebeck G, et~al (2009) Detecting spam in a twitter
  network. First Monday 15(1)

\bibitem[{Zinkevich(2003)}]{zinkevich2003online}
Zinkevich M (2003) Online convex programming and generalized infinitesimal
  gradient ascent

\end{thebibliography}


\end{document}